\documentclass[nohyper,11pt,letterpaper]{JHEP3}
\usepackage{epsfig}

\newcommand{\be}{\begin{equation}}
\newcommand{\ee}{\end{equation}}

\author{Dejan Stojkovic \\
       MCTP, Department of Physics, University of Michigan,\\
       Ann Arbor, MI 48109-1120 USA\\
      E-mail: \email{dejans@umich.edu}
}

\abstract{We study various configurations in which a domain wall (or cosmic string),
described by the Nambu-Goto action, is embedded in a background space-time
of a black hole in $(3+1)$ and higher dimensional models.  We calculate
 energy fluxes through the black hole horizon. In the
simplest case, when a static domain wall enters  the horizon of a static
black hole perperdicularly, the energy flux is zero. In more complicated
situations, where  parameters which describe the domain wall surface
are time and position dependent, the flux
is non-vanishing is principle. These results are of importance in various
conventional  cosmological models which accommodate the existence of
domain walls and strings and also in brane world scenarios.
}


\preprint{ MCTP-04-50
}

\title{Energy flux through the horizon in the black hole-domain wall
systems}

\begin{document}

\section{Introduction}

Topological defects can arise in a wide class of
cosmological models. In principle, classical field theory models, embedded
in a particular cosmological model, which admit non-trivial topology
give rise to topological defects. Most of grand unifying
theories (GUT) \cite{ViSh} and some extensions of the electroweak standard
model \cite{ssv} support the existence of topological defects.
This is the primary motivation for the study of topological defects in the
early universe.

According to the standard cosmology, evolution of domain walls in the
universe is such that they quickly come to dominate the energy
density of the universe. Such domination would severely
violate many observational astrophysical constraints, so that domain walls
represent a cosmological disaster. If the universe produces domain walls,
it must get rid of them before the nucleosynthesis epoch at the latest.

In  \cite{SF}, a simple solution to the cosmological monopole problem was
proposed. Primordial black holes,
produced in the early universe, can accrete magnetic monopoles within
the horizon before the relics  dominate the energy density of the
universe. One could hope that a similar idea can be applied to
the domain wall problem. However, the extended nature of domain wall
topological defects (in particular its super-horizon size) makes the
domian wall problem much more difficult to treat.

The question of how black holes interact with domain walls is not
trivial. For example, if a static, planar domain wall enters a static
black hole perpendicularly to the horizon (i.e. the null Killing vector
which generates the horizon hypersurface is tangent to the domain wall
surface), the energy flux through the horizon is zero due to the
symmetries of the system. In this idealized case, the black hole cannot
accrete energy from the wall. However, the situation is different in more
complicated cases. In a more realistic case, we expect domain wall to have
some velocity with respect to the black hole. As the domain wall
encounters the black hole, its surface will deform and parameters
describing the domain wall surface will be time and/or position dependent.
In section \ref{sec:sch} we will show that in this case the energy that
crosses the horizon is non-vanishing.

 In section \ref{sec:rot} we study the case where a black hole is
rotating. A rotating black hole can also accrete the energy from the
domain wall. However, the term proportional to the rotational parameter
has an opposite sign from a leading order term which signals that rotation
works in the opposite direction --- it helps extraction of energy from the
black hole and reduces acreetion.

Another important reason to study this question  comes from
so-called brane world models with large extra dimensions \cite{ADD}.
In this framework our universe is just a three dimensional domain
wall (a brane) embedded in a higher dimensional space. All
the standard model particles are localized on the brane while gravity
can propagate everywhere. In particular, black holes being
gravitational solitons can propagate in a higher dimensional bulk
space. In the simplest formulation, the gravitational field of the brane
is neglected and extra dimensions are flat. An important generic feature
of this model is that the fundamental quantum gravity mass scale
$M_*$ may be very low  (of order  TeV) and the size of
the extra spatial
dimensions may be much larger than the Planck length ($\sim
10^{-33}$cm). The maximal size $L$
of  extra spatial dimensions, allowed by the experiments testing the
deviations from Newton's law at shortly distances, is the order of
0.1 mm. The gravitational radius $R_0$ of a black
hole of mass $M$ in the spacetime with $k$ extra dimensions is defined
by the relation $G^{(4+k)}\, M\sim R_0^{k+1}$, where
$G^{(4+k)}=1/M_*^{(k+2)}$ is the $(4+k)$-dimensional Newton coupling
constant. The minimal mass of the black hole is determined by the
condition that its gravitational radius coincides with its Compton
length $\sim 1/M$. The mass of such an elementary black hole is $M_*$.
For $M_*\sim TeV$ one has $R_*\sim 10^{-17}$cm. When $M\gg M_*$ the
higher dimensional mini black holes can be described by the classical
solutions of vacuum Einstein's equations. It is
assumed that the size of a black hole $R_0$ is much smaller than the
characteristic size of extra dimensions, $L$,  and neglect the effects of
the black hole deformation connected with this size.

In this framework, there exist interesting
possibility of production of mini black holes in future  collider
and cosmic rays experiments. Estimates \cite{BHacc} indicate that the
probability for creation of a mini black hole in near future hadron
colliders  such as the LHC (Large Hadron Collider)  is so high that they
can  be called  ``black hole factories".
After the black hole is formed  it decays by emitting Hawking radiation.
As a result of the emission of the graviton into the bulk space, the black
hole recoil can move the black hole out of the brane \cite{FrSt,St}. Black
 hole radiation
would be terminated and an observer located on the brane would
register  virtual energy non-conservation.
 In order to quantify this effect, it is important to know more details
about interaction of the black hole with the brane. In particular, the
difference in energy of the black hole before and after it leaves the
brane (the energy "cost" of leaving the brane) will strongly depend on the
amount of energy which crosses the horizon in the process of this
time-dependent interaction. This question is analyzed in section
\ref{sec:brane}.

In section \ref{sec:string} we analyzed the case of a cosmic string.

The interaction of various topological defects
with black holes has been studied before. In \cite{FrolovBHS} and
\cite{FrolovBHDW} interaction of black holes with cosmic strings and
domain walls in $(3+1)$-dimensional universe was studied. This study
was generalized in \cite{FSS}  to the case
where both the brane and the bulk space in which the brane is moving may
have an arbitrary number of dimensions. The relevant calculation was done
in the weak filed approximation where the  black hole is far from the
brane. Rotating black holes were studied in the limit of slow
rotation. The question of energy flux trough the horizon in the black
hole-defect system was not studied by now.

\section{Schwarzschild black hole in $3+1$ dimensions }
\label{sec:sch}

We consider an axially symmetric domain wall in a background of
 Schwarzschild black hole in $(3+1)$-dimensions. The Schwarzschild
 background metric in standard coordinates $(t,r,\theta,\phi)$ is:

\be
ds^2= -(1-\frac{2GM}{r}) dt^2 + (1-\frac{2GM}{r})^{-1} dr^2 + r^2
d\theta^2  + r^2 \sin^2(\theta) d\phi^2  \, ,
\ee
where $M$ is the mass of the black hole and $G$ is the Newton's
gravitational constant.

In vicinity of the horizon ($r_h=2GM$), it is convenient to work in
a new set of coordinates $(u,r,\theta,\phi)$, where
new timelike coordinate is defined as:

\be
u= t+r + 2GM \ln \left( \frac{r}{2GM} -1 \right)  \, .
\ee
This change yields:
\be \label{sw4d}
ds^2= -(1-\frac{2GM}{r}) du^2 + 2du dr + r^2 d\theta^2 + r^2
\sin^2(\theta)  d\phi^2  \, .
\ee

\begin{figure}[tbp]
\centerline{\epsfxsize = 0.40 \hsize \epsfbox{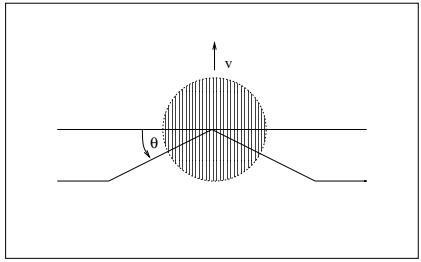}}
\caption{A black hole encountering the domain wall with a
relative velocity $v$. } \label{thetadot}
\end{figure}

Induced metric on a domain wall world-sheet $\gamma_{ab}$, in a given
 background metric $g_{\mu \nu}$ is:

\be
\gamma_{ab}= g_{\mu \nu} \partial_a X^{\mu} \partial_b X^{\nu} \, ,
\ee
where Latin indices go over internal domain wall world-sheet coordinates
 $\zeta^a$, while Greek indices go over
space-time coordinates, $X^{\mu}$. We fix the gauge freedom due to
world-sheet coordinate reparametrization by choice:

\be \label{gauge}
 \zeta^0 = X^0 = u, \ \  \zeta^1 = X^1  = r , \ \
 \zeta^2 =  X^2 =\phi \ .
\ee
The remaining coordinate $\theta$  describes the motion of
 a domain wall surface in the background space-time. If $\theta =$const,
i.e. domain wall is static, one could argue that due to symmetries of the
system the energy flux through the horizon is zero. However, the situation
is different if for example $\theta = \theta (u,r)$.   This can happen
when a domain wall encounters the black hole with some relative
velocity $v$ (see Fig. \ref{thetadot}).

With this choice, we can calculate the non-zero elements of
the induced metric:
\begin{eqnarray}
\gamma_{uu} &=& -(1-\frac{2GM}{r}) + r^2 {\dot \theta}^2
\\
\gamma_{rr} &=&  r^2 \theta'^2
\nonumber \\
\gamma_{\phi \phi } &=&  r^2 \sin^2(\theta )
\nonumber \\
\gamma_{ur} &=& 1+ r^2 \theta' {\dot \theta} \, .
\nonumber
\end{eqnarray}
Here, a dot and a prime denote derivation with respect to the coordinates
$u$ and $r$ respectively.

The dynamics of the domain wall can be derived
from the Nambu-Goto action:

\be \label{Sng}
S= -\sigma \int \sqrt{-\gamma} d^3 \zeta  \, ,
\ee
where $\sigma$ is the energy density of a domain wall and $\gamma$ is the
 determinant of the induced metric
$\gamma_{ab}$.

From the Killing equation and the conservation of momentum-energy
tensor, $T^{\mu \nu}$ it follows that  whenever there is a symmetry of the
system (described by a Killing vector $\xi^\mu$ for a given geometry)
there is a covariantly conserved quantity $T^{\mu \nu}\xi_\nu$.   If the
Killing vector is $\xi_{(u)}^{\mu}$, describing
invariance with respect to translations in time, the conserved charge is
energy. Then, the vector $T^{\mu \nu} \xi_{(u) \mu} $
%= T^\mu_{\, 0}
%=-T^{\mu 0}$ (for the metric for which $g_{00}=-1$ at infinity),
 can be
interpreted as the negative of the energy flux through some hypersurface
as seen by an observer at infinity.
%The corresponding
%conserved charge is then defined as
%\be
% Q= \int T^{\mu \nu} \xi_{\mu} d \sigma_{\nu} \, ,
%\ee
%where integration goes over some hypersurface whose element is $d
%\sigma_{\nu}$.
Thus, the total energy which passes through some hypersurface whose
element is $d \sigma_{\nu}$ is

 \be \label{E} \Delta E = -\int T^{\mu \nu}
\xi_{(u) \mu} d \sigma_{\nu} \, .
\ee
Similar result is valid for the total angular momentum

\be\label{J}
\Delta J = \int T^{\mu \nu} \xi_{(\phi)
\mu} d \sigma_{\nu}  \, .
\ee

Here, $\xi_{\mu}$ is the Killing vector  for a given geometry. In
particular  $\xi_{(u)}^{\mu}  = \delta^{\mu}_{u}$ and $\xi_{(\phi)}^{\mu}
= \delta^{\mu}_{\phi}$. Since we are interested in behavior
near the horizon of the  black hole, we take $d \sigma_{\nu}$
to be the element of horizon null-hypersurafce:

\be
d \sigma_{\nu} = \sqrt{-g} \delta^{r}_{\nu} du d\theta d\phi \, .
\ee
Here, we took that the unit vector defining the horizon null-hypersurafce
points outwards from the black hole.
 Thus, we get:

\be \label{Tur}
\Delta E = -\int \sqrt{-g} T_u^r  du d\theta d\phi  \, .
\ee
This expression gives the change of energy of the domain wall. The
corresponding change of energy of the black hole will be the same in
magnitude with an opposite sign. From now on we discuss changes in
quantities defined for a black hole.

For the metric (\ref{sw4d}) we have
$T_u^r = g_{uu}T^{ur}+g_{ur}T^{rr}$. At the horizon, the metric component
$g_{uu}$ vanishes so the only contribution is from $T^{rr}$. From the
action (\ref{Sng}) we can derive the momentum-energy tensor:

\be \label{MET}
\sqrt{-g} T^{\mu \nu} = -\sigma \int \delta^4
\left[ X^\mu - X^\mu (\zeta)   \right]
\sqrt{-\gamma} \gamma^{ab} \partial_a X^{\mu} \partial_b X^{\nu} d^3 \zeta
\, . \ee

%Substituting this into (\ref{Tur}) we get:

%\be
%\Delta E = \sigma \int \delta
%\left[ \theta - \Theta (u,r)   \right] \sqrt{-\gamma} \gamma^{rr}  du
 %d\theta d\phi  \, .
%\ee

The final expression for the energy which crosses the horizon of the black
hole ($r_h=2GM$) is  thus:
\be\label{finalEsw}
\Delta E = \sigma 16 \pi G^3M^3 \int \delta
\left[ \theta - \Theta (u,r)   \right]
\frac{{\dot \theta}^2 \sin (\theta) }{\sqrt{1+8G^2M^2{\dot \theta}
\theta'}}  du d\theta \, .
\ee

The second order equations of motion for $\theta (u,r)$ which follow from
 the action
(\ref{Sng}) are rather complicated and we are not going to present them
 here. One can check that they do not give any constraints on ${\dot
 \theta}$ and $\theta'$.

Canonical momentum, $P_{\mu} = \frac{\partial L}{\partial {\dot X}^\mu}$
 can be derived from the action (\ref{Sng}):

\be
P_{\mu} = - \sqrt{-\gamma} \gamma^{u b} \partial_b X^\mu \, .
\ee
This form of momentum, in general case, gives three constraints on
 dynamical variables:

\be
P_\mu \partial_i X^\mu =0
\ee
and
\be
P_\mu P^\mu + h =0 \, ,
\ee
where index $i$ goes over spatial indices $r, \phi$,
while $h$ is the determinant of the spatial part of the metric
 $\gamma_{ab}$. With the gauge choice (\ref{gauge}), these constraints are
automatically satisfied and do not give any further constraints on ${\dot
\theta}$ and $\theta'$.

For simplicity, let us analyse expression (\ref{finalEsw}) under
assumption that $\theta$ does not change much with radial
coordinate $r$, i.e. $\theta ' \approx 0$. In that case we have

\be \label{finalEsw1}
\Delta E =\sigma 16 \pi G^3M^3 \int \delta
\left[ \theta - \Theta (u)   \right] {\dot \theta}^2 \sin (\theta)  du
d\theta \, .
\ee

The overall sign of $\Delta E$ is positive signaling that the energy is
flowing into the black hole, i.e. the black hole grows. In other words,
domain wall gets ``eaten" by the black hole.
Strictly speaking, eq. (\ref{finalEsw1}) gives the total energy which
crosses the horizon since $T^{\mu \nu}$ in (\ref{MET}) is the total
momentum-energy tensor of the domain wall containing both kinetic
and ``static" energy of the wall.

The amount of energy which
crosses the horizon obviously depends on ${\dot \theta}$. If the process
is slow and adiabatic (quasi-static), then ${\dot \theta} \rightarrow 0$.
This implies $\Delta E \rightarrow 0$, which means that not a significant
amount of energy crosses the horizon.

The case when ${\dot \theta}$ is large is more interesting. This can
happen, for example, when the black hole encounters the domain wall with
large relative velocity $v$, say $v \approx c$. In that case $R_{bh}
\theta \approx v u$. $R_{bh}$ here is the horizon
radius ($r_h$) of the black hole. Thus, we have ${\dot \theta} \approx
v/R_{bh}$. We assume that energy is flowing through the horizon
while the angle $\theta$ is taking values from $\pi/2$ to $\pi$.
Performing the integration in (\ref{finalEsw}) and taking  $\int
\sin \left[ \theta (u) \right] du \approx R_{bh}/v
\int_{\frac{\pi}{2}}^\pi \sin (\theta)  d\theta = R_{bh}/v $ we have

\be \label{Esch}
\Delta E  \approx  \sigma R_{bh}^2 v \, ,
\ee
where we dropped factors of order of unity.
This result agrees with the one that could be guessed on dimensional
grounds.

Note that the expression for energy in (\ref{finalEsw1}) depends only on
${\dot \theta}^2$, i.e it does not change the sign when  ${\dot
\theta}$ changes the sign. If the domain wall oscillates, then there will
be several cycles and in each of them the energy which crosses horizon
will be of order given in (\ref{Esch}).

For completeness, let us mention that the angular momentum flux through
the horizon of the Schwarzschild black hole is zero as expected.

 \section{Rotating black hole in $3+1$ dimensions}
\label{sec:rot}

We now calculate the energy  flux through the
horizon of a rotating black hole. In \cite{FSS},  a domain wall in the
 background of a rotating black hole
was studied. Angular momentum fluxes corresponding to various positions of
a domain wall (i.e. in equatorial and azimuthal planes etc.) were
calculated. Since the solution describing the shape of the wall was
stationary (time independent) no energy flux through the horizon was
found. Here, we extend this analysis to the case where the parameters
describing the wall's world sheet can be time dependent.

 We consider  an axially symmetric
domain wall in a background of  a rotating black hole in $3+1$ dimensions.
The rotating black hole  background in  coordinates $(u,r,\theta,\phi)$
is:

\begin{eqnarray} \label{kerr4d}
ds^2 &=& -(1-\frac{2GMr}{\rho^2}) du^2 +2du dr + \rho^2 d\theta^2
-2a\sin^2(\theta) drd\phi \\ &+&  \frac{ (r^2+a^2)^2 - \Delta a^2
\sin^2(\theta) }{\rho^{2}} \sin^2(\theta) d\phi^2 -\frac{4aGMr}{\rho^2}
dud\phi \nonumber
\end{eqnarray}
where $a$ is the rotational parameter and
\begin{eqnarray}
&& \rho^2 =r^2+a^2\cos^2(\theta) \\
&& \Delta =r^2-2GMr +a^2  \nonumber \, .
\end{eqnarray}

The event horizon is given by the solution of the equation $\Delta =0$,
i.e.

\be
r_h = GM + \sqrt{G^2M^2 -a^2}
\ee
The static limit surface (or the infinite redshift surface) is given by
the solution of the equation $g_{00}=0$, i.e. $r_{sl} = GM + \sqrt{G^2M^2
-a^2\cos^2(\theta)}$, and its position does not coincide with the horizon.
The region between the horizon and the static limit surface is known as
"ergosphere".

We fix the gauge freedom due to domain wall world-sheet
coordinate reparametrization similarly as before:

\be
 \zeta^0 = X^0 = u, \ \  \zeta^1 = X^1  = r , \ \
 \zeta^2 =  X^2 =\phi \ .
\ee
We consider the case where the remaining coordinate $\theta$
which describes the motion of  a domain wall surface is
$\theta =\theta (u,r)$. With this choice, the non-zero elements of
the induced metric are:

\begin{eqnarray}
\gamma_{uu} &=& -1+\frac{2GMr}{\rho^2}+\rho^2 \dot{\theta}^2
\\
\gamma_{ur} &=& 1 + \rho^2 \dot{\theta} \theta' \nonumber \\
\gamma_{u \phi} &=& -\frac{2 aGMr \sin (\theta)^2}{\rho^2} \nonumber \\
\gamma_{rr} &=& \rho^2 \theta'^2 \nonumber \\
\gamma_{r \phi } &=&  -a  \sin^2(\theta)
\nonumber \\
\gamma_{\phi \phi } &=& (r^2+a^2)\sin^2(\theta)  +\frac{2a^2GMr
\sin^4 (\theta) }{\rho^2} \, . \nonumber
\end{eqnarray}

The energy  flux can be calculated from eq.
 (\ref{E}). The relevant Killing vector for the metric (\ref{kerr4d})
 is  again $\xi_{(u)}^{\mu}  = \delta^{\mu}_{u}$.

From (\ref{kerr4d}) we also have have $T_u^r =
g_{uu}T^{ur}+g_{ur}T^{rr}+g_{u\phi}T^{\phi r}$.
%and $T_\phi^r =g_{u\phi}T^{ur}+g_{\phi \phi}T^{\phi r}+g_{\phi r}T^{rr}$.
The corresponding energy  flux at the horizon
($r_h = GM + \sqrt{G^2M^2 -a^2}$) is non-vanishing:

\be \label{Ehor}
\Delta E =  4\pi \sqrt{2} \sigma G^3M^3 K \int \delta\left[ \theta -
\Theta (u,r) \right] \frac{ \dot{\theta}^2 \sin (\theta)
\sqrt{K-\frac{1}{2} \left(\frac{a}{GM} \right)^2\sin^2(\theta) } }
{\sqrt{1+\dot{\theta}^2a^2 \sin^2(\theta) +4\dot{\theta} \theta'G^2M^2K} }
du d\theta   \, , \ee
where $K=1+\sqrt{1-\left(\frac{a}{GM} \right)^2}$.
For  $a=0$ we recover the result (\ref{finalEsw}) for a non-rotating black
hole. For an extremal black hole $a=GM$, eq. (\ref{Ehor}) becomes
\be
\Delta E_{\rm extremal} =  4\pi  \sigma G^3M^3  \int \delta\left[
\theta - \Theta (u,r) \right] \frac{ \dot{\theta}^2 \sin (\theta) (1+\cos^2
\theta ) }
{\sqrt{1+\dot{\theta}^2G^2M^2 \sin^2(\theta) +4\dot{\theta} \theta'G^2M^2}
} du d\theta   \, . \ee

In order to illustrate some interesting consequences, we expand
(\ref{Ehor}) in terms of small $\dot{\theta}$, $\theta'$ and $a$, and keep
only the leading order terms. We get

\be
\frac{d E}{du d\theta } =  16\pi \sigma \sin (\theta) G^3M^3
\dot{\theta}^2 -  64 \pi \sin (\theta) G^5M^5 \dot{\theta}^3 \theta'
- 2\pi \sin (\theta) GM  (4-\cos^2(\theta) ) a^2
\dot{\theta}^2  \, .
 \ee
The term which depends on rotational parameter $a$ has an opposite sign
from the leading order term. As expected, rotation of the black hole
helps extraction of energy from the black hole and reduces acreetion.

 \section{Higher dimensional static black hole }
\label{sec:brane}

Higher dimensional black holes are of particular interest in theories with
large extra dimensions. In these models, all the standard model fields are
localized on the $(3+1)$-dimensional brane while geometrical degrees of
freedom can propagate everywhere. This implies that small black holes can
leave our brane. The probability for something like this to happen due
to Hawking radiation was studied in \cite{FrSt}. The difference in energy
between the configuration where the black hole is on the brane and the one
where the black hole is in the bulk was not calculated, it was only
estimated on dimensional grounds. Here, we calculate the energy flux
through the horizon during the process of the black hole extraction from
the brane. In the rest frame of the black hole this situation corresponds
to the interaction with a non-static brane.

We consider a non-rotating higher-dimensional  black hole which is a
simple generalization of the Schwarzschild solution in $(3+1)$-dimensional
space-time.. Although it is
straightforward to write down the metric in an arbitrary number of
dimensions, all the basic results can be presented in $(4+1)$-dimensional
space-time.

The metric, in coordinates $(u,r,\theta,\phi, \psi)$, is
  \be
\label{sw5d} ds^2= -(1-\frac{R_0^2}{r^2}) du^2 + 2du dr + r^2 d\theta^2 +
r^2 \sin^2(\theta) d\phi^2 + r^2 \cos^2(\theta) d\psi^2 \, ,
\ee
where $R_0$ is the gravitational radius of the $(4+1)$-dimensional black
hole. The extra angular variable $\psi$ takes values from the interval
$[0,2\pi]$.

The domain wall, which represents our universe is $(3+1)$-dimensional.
We fix the gauge freedom due to domain wall world-sheet
coordinate reparametrization in this way:

\be
 \zeta^0 = X^0 = u, \ \  \zeta^1 = X^1  = r , \ \
\zeta^2 =  X^2 =\theta \ \ \zeta^3 =  X^3 =\phi \ .
\ee

Let us  consider the case where the remaining coordinate $\psi$
which describes the motion of  a domain wall surface is a
function of time and radial coordinate, ie. $\psi  =\psi (u,r)$. With this
choice, the non-zero elements of the induced metric are:

\begin{eqnarray}
\gamma_{uu} &=& -(1-\frac{R_0^2}{r^2}) + r^2 \cos^2(\theta){\dot \psi}^2
\\
\gamma_{rr} &=&  r^2 \cos^2(\theta)  \psi'^2 \nonumber  \\
\gamma_{\phi \phi } &=&  r^2 \sin^2(\theta)
\nonumber \\
\gamma_{\theta \theta } &=&  r^2
\nonumber \\
\gamma_{ur} &=& 1 + r^2 \cos^2(\theta){\dot \psi}  \psi' \, . \nonumber
\end{eqnarray}

This yields the expression for the energy which crosses the horizon
($r_h=R_0$)

\be
\Delta E =  \sigma R_0^4 \int \delta
\left[ \psi - \Psi (u,r)   \right] \frac{ {\dot \psi}^2 \sin (\theta)
\cos^2(\theta) }{ \sqrt{1+ 2\cos^2(\theta) R_0^2 {\dot \psi} \psi' } }
du d\theta d\phi d\psi  \, . \ee

Note that the brane tension (energy density) $\sigma$ now has dimensions
of (mass)$^4$. If for simplicity we set $ \psi' \approx 0$ and integrate
over angular variables $\theta$ and $\phi$ we get

\be \label{finalEsw5d}
\Delta E =  \frac{2}{3} \pi \sigma R_0^4 \int \delta
\left[ \psi - \Psi (u)   \right]
{\dot \psi}^2  du  d\psi \, .
\ee

Here again, the sign signals that the net energy is flowing toward the
black hole. We can use a similar approximation as earlier, $R_0 \psi
\approx v u$, where $v$ is the relative black hole-brane velocity. After
integration we get

\be \label{Ebrane}
\Delta E  \approx \sigma R_0^3 v\, ,
\ee
where we dropped the terms of order unity.
This expression agrees with the one used in \cite{FrSt}.

\section{Cosmic string}
\label{sec:string}

We now consider the case of a cosmic string described by a Nambu Goto
action in the background metric of a Schwarzschild black hole in $3+1$
dimensions.
The background metric is given in (\ref{sw4d}).
We fix the gauge freedom due to string world-sheet
coordinate reparametrization in this way:

\be
 \zeta^0 = X^0 = u, \ \  \zeta^1 = X^1  = r \ .
\ee

The string world sheet is fully specified with two coordinates $(u,r)$,
while the background is $(3+1)$-dimensional $(u,r,\theta,\phi)$. Thus, we
have two embeding coordinates $\theta$ and $\phi$. Let us  consider the
case where the two remaining coordinates $\theta$ and $\phi$ which
describe the motion of  a string in a background speace time are functions
of time and radial coordinate, ie. $\theta =\theta(u,r)$ and $\phi =\phi
(u,r)$. With this choice, the non-zero elements of the induced string
metric are:

\begin{eqnarray}
\gamma_{uu} &=& -(1-\frac{2GM}{r}) + r^2 {\dot \theta}^2+ r^2
\sin^2(\theta){\dot \phi}^2 \\
\gamma_{ur} &=& 1+ r^2 {\dot \theta}\theta'+ r^2
\sin^2(\theta){\dot \phi}\phi'
\nonumber \\
\gamma_{rr} &=& r^2 \theta'^2+r^2 \sin^2(\theta)  \phi'^2 \, .
\nonumber \end{eqnarray}

This yields the expression for the energy which crosses the horizon
($r_h=2GM$) of the black hole

\be
\Delta E =  \int  \frac{ \mu 4 G^2M^2 \delta
\left[ \theta - \theta (u,r)   \right] \delta
\left[ \phi - \Phi (u,r)   \right] \left( {\dot \theta}^2 +\sin^2(\theta)
{\dot \phi}^2 \right) du d\theta d\phi  }{ \sqrt{1 + 8G^2M^2 \left( {\dot
\theta}\theta' -  \sin^2(\theta) {\dot \phi}\phi' \right)
-\sin^2(\theta) 16G^4M^4  \left( {\dot
\theta}^2\phi'^2  + \theta'^2 \dot{\phi}^2  +  2{\dot
\theta}\theta' {\dot \phi}\phi'  \right)
  } } \, . \ee
Here, $\mu$ is the energy density per unit length of the string.
%It is interesting that the angular momentum flux through the horiozn is
%also non-vanishing:

%\be
%\Delta J =  \int  \frac{ - \sigma 4 G^2M^2 \delta
%\left[ \theta - \theta (u,r)   \right] \delta
%\left[ \phi - \Phi (u,r)   \right] \sin^2(\theta) \left[ {\dot \phi}
%+4G^2M^2\left(    {\dot \phi} {\dot \theta}\theta' - \phi'{\dot \theta}^2
%\right) \right] du d\theta d\phi  }{ \sqrt{1 + 8G^2M^2{\dot
%\theta}\theta'-\sin^2(\theta) 16G^4M^4  \left( {\dot \theta}^2\phi'^2  +
%\theta'^2 \dot{\phi}^2 \right) -  \sin^2(\theta) 8G^2M^2{\dot \phi}\phi'
%\left(1+ 4G^2M^2{\dot \theta}\theta' \right)   } }    \, . \ee
%This is the consequence of the fact that the coordinate $\phi$ is a
%function of time and position. Since in a generic case (with no special
%symmetries) we expect both $theta$ and $\phi$ to be time and position
%dependent, a non-rotating black hole would gain some angular momentum
%in an encounter with a cosmic string.

If we fix the coordinate $\phi$, say $\phi =0$, we get

\be
\Delta E = \mu 8\pi G^2M^2 \int \delta
\left[ \theta - \theta (u,r)   \right] \frac{   {\dot \theta}^2    }{ \sqrt{1 + 8G^2M^2{\dot
\theta}\theta'  } }   du d\theta   \, . \ee
This is just an analog of the result (\ref{finalEsw}) for domain walls. In
order to quantify the effect, we can apply a similar approximation of a
large relative velocity $v \approx c$. In this case we get
$ \Delta E \approx \mu R_{bh}$.

%The corresponding angular momentum flux is zero in this case.

The other configurations, like a cosmic string in the background
space-time of a rotating (and/or higher dimensional) black hole, can be
treated in analogy with domain wall treatments.

\section{Conclusions}

We addressed the question of the energy flux through the horizon during
black hole interaction with domain walls and strings. In the simplest
case, when a static domain wall (or string) enters  the horizon of a
black hole perperdicularly the energy flux is zero. In more complicated
situations, the flux could be non-vanishing. For example, if one of the
parameters which describes the domain wall surface in the black hole
background is time and position dependent, the net flux through the
horizon is non-zero. These results are of importance in
various cosmological models which accommodate the existence of domain
walls and strings. In particular, the cosmological domain wall problem
could be alleviated if primordial black holes can accrete a significant
portion of the energy density contained in walls (we note however, that
the extended (super-horizon) nature of domain walls requires a very
careful treatment). Another possible framework of interest could be
interaction of large black holes located in centers of galaxies with
domain walls and strings.

A similar analysis can be done in the framework of brane world scenarios.
Going from $(3+1)$-dimensional models to higher dimensional ones, only
quantitative results change. The qualitative picture remains the same.
These results are of importance in studying the interaction of our world
with small higher dimensional black holes. In particular, a
probability for a black hole to go off the brane in various processes
will strongly depend on the amount of energy which crosses the horizon
during the process of the extraction from the brane.

There are some model independent features for the various scenarios we
considered. As expected, there is the net energy flow through the horizon
only if the configuration in question is time dependent. For quasi-static
processes the amount of energy which crosses the horizon is
negligible. Otherwise, the energy is always proportional to the area
(volume) of  the domain wall-brane cross section.

\vspace{12pt} {\bf Acknowledgments}:\ \
The author is grateful to  Glenn Starkman, Jim Liu
and especially Valeri Frolov for very useful discussions.
The author was suppoted by DOE at the Michigan Center for
Theoretical Physics, University of Michigan.

\end{document}